\documentclass[aps,prc,preprint,groupedaddress]{revtex4-1}

\usepackage{graphicx}
\usepackage{comment}

\begin{document}

\title{Transverse mass spectra and scaling of hadrons at RHIC and LHC energies}

\author{\large P. K. Khandai$^1$}
\author{\large P. Sett$^{2,3}$}
\author{\large P. Shukla$^{2,3}$}
\email{pshukla@barc.gov.in}
\author{\large V. Singh$^1$}

\affiliation{$^1$Department of Physics, Banaras Hindu University, Varanasi 221005, India}
\affiliation{$^2$Nuclear Physics Division, Bhabha Atomic Research Center, Mumbai 400085, India}
\affiliation{$^3$Homi Bhabha National Institute, Anushaktinagar, Mumbai 400094, India}


\begin{abstract}
  We present a systematic study of transverse mass ($m_T$) spectra of mesons and baryons at 
RHIC and LHC energies. In an earlier study, it was shown that all the mesons produced in 
p+p and d+Au collisions at $\sqrt{s_{NN}}$ = 200 GeV follow $m_T$ scaling while in Au+Au collisions 
at same energy the mesons with strange and charm quark contents do not follow $m_T$ scaling which
can be attributed to medium modifications.
  We extend this study for baryons produced in all above colliding systems at 
$\sqrt{s_{NN}}$ = 200 GeV. Although all available baryon spectra behave differently from 
mesons but are found to scale with protons for both p+p and d+Au collisions. In case of 
Au+Au collisions, the strange baryons behave differently from protons. 
  This study has also been performed on the meson spectra in p+p collisions
at 62.4 GeV which lies in between the highest RHIC energy and SPS energy where the $m_T$ 
scaling was first observed and we arrive at same conclusions as those at 200 GeV. 
  We test the $m_T$ scaling at LHC with the first meson and baryon spectra measured in p+p collisions at  
900 GeV and find that the $m_T$ spectra of kaons and $\phi$ do not scale with pions that well.
Here we use all data measured in the mid rapidity region. 
Proton to pion ratio is studied as a function of $m_T$, at different energies and for different systems and is found
that this ratio for p+p at 900 GeV is quite low from that in p+p collisions at RHIC energies.

\vspace{0.6in}

\begin{center}
Submitted to Physical Review C
\end{center}

\end{abstract}



\maketitle

\section{Introduction}
  Heavy ion collisions at relativistic energies are performed to study the properties of 
matter at high temperature where a phase transition to Quark Gluon Plasma (QGP) is expected.
  In Au+Au collisions at $\sqrt{s_{NN}}$ = 200 GeV at Relativistic Heavy Ion Collider (RHIC), 
many signals point to the formation of Quark Gluon Plasma (QGP) \cite{WP,HADPHEN,PHOPHEN,DIELEC}. 
RHIC continues to study the detailed properties of the strongly interacting matter using p+p, d+Au and Au+Au 
systems at various colliding energies from 200 GeV down up to 7.7 GeV.
 First results from Pb+Pb collisions at LHC already giving possible glimpses of QGP from 
precise measurements of observables such as jets \cite{CMSJET} and 
quarkonia \cite{CMSUP,CMSQUARK}. First measured charged particle spectra from LHC \cite{LHC} are 
also available which give the global features of the collisions.

   Measurements of transverse momentum spectra for particles emerging from p+p collisions 
are used as a baseline to which similar measurements from heavy ion collisions are compared
to study collective effects.
 Long ago WA80 collaboration found that the spectral shapes of $\pi$ and $\eta$ mesons 
measured in S + S (200 AGeV) collisions  are identical when plotted as a function of $m_T$ \cite{WA80}. 
This property is known as $m_T$ scaling and has been extremely useful to extrapolate the 
meson spectra to unkown regions and to obtain the mesons spectra which are not measured. 
  One application of such analysis is to obtain cocktail of decay products from produced 
hadrons \cite{DIELEC,COCK}. Obtaining the integrated yields of hadrons is another important application   
where $m_T$ scaling can be used.

  In an earlier study \cite{mesonscaling}, we found that all mesons follow $m_T$ scaling in 
p+p and d+Au collision at $\sqrt{s_{NN}}$ = 200 GeV, while the mesons with strange and charm quark 
contents do not follow it in Au+Au collisions at the same energy which can be associated to 
medium modifications. The more  detailed centrality analysis of Au+Au collision reveal that 
in case of $\eta$, the data of all centralities are very well reproduced by $m_T$ scaled pion data.
The ratios obtained also remain approximately same for all centralities. For particles, kaon and $\phi$,
peripheral collision data is better reproduced as compared to central collision data and 
the ratios also increase as the collision becomes more central.  
  We extend this study for baryons produced in all above colliding systems at 
$\sqrt{s_{NN}}$ = 200 GeV.  We study proton and strange baryons such as 
$\Lambda$, $\Xi$, $\Omega$ and their anti-particles.
Although all available baryon spectra behave differently from 
mesons but are found to scale with protons for both p+p and d+Au collisions. 
In case of Au+Au collisions, the strange baryons behave differently from protons. 
  
  This study has also been performed on the meson spectra in p+p collisions
at 62.4 GeV. This lies in between the SPS energy and highest RHIC energy. 
The conclusions at 200 GeV hold at 62.4 GeV. 
 Finally and most importantly we test the $m_T$ scaling at LHC with the first 
meson and baryon spectra measured in p+p collisions at 900 GeV. It is found that 
the kaons and the $\phi$ do not scale that well with pions; but in case of baryon,
the $m_T$ scaled curve is in good agreement with the $\Lambda$ and with the small data set of $\Xi$.  
 We also study the proton to pion ratio in different systems as a function of energy.
It is found that this ratio for p+p at 900 GeV is quite low from that in p+p collisions 
at RHIC energies.

 \section{Fit procedure using $m_{T}$ scaling}
 In this section, we describe the procedure of fitting of hadron spectra using $m_{T}$ scaling.
 To give a good descritption of pion spectra in wide $p_T$ range, the PHENIX collaboration 
\cite{PPG077, PPG099, SEANA, DIELEC} used a simple form referred as the modified Hagedorn 
formula. This formula has been extensively used for p+p, d+Au and Au+Au collisions \cite{mesonscaling}
which in terms of $m_T$ is given by: 
  \begin{eqnarray}\label{fitfun}
 E{d^3N \over dp^3} & = & \frac {A} { \left[ {\rm exp}(- a m_{T} - b m_{T}^2) 
                             + {m_{T} \over p_0} \right]^n },  \nonumber \\ 
      & =  & f_{\pi/p} \left( \sqrt{p_T^2 + m_{\pi/p}^2} \right),
\end{eqnarray}
which is close to an exponential form at low $p_{T}$ and a pure power law form at high $p_{T}$.
Here $f_{\pi/p}$ is the pion or proton fit function, $m_{\pi/p}$ is the rest mass of pion or proton
and $A$, $a$, $b$, $p_0$ and $n$ are the fit parameters. 
Then we obtain the spectra of mesons (or baryons) using pion (or proton) fit function as: 
\begin{eqnarray}
 E{d^3N \over dp^3}  =  S \,\, f_{\pi/p} \left( \sqrt{p_T^2 + m_{h}^2} \right),
\end{eqnarray}
where  $m_{h}$ is the rest mass of the corresponding meson (or baryon). 
The factor $S$ is the relative normalization of the meson (or baryon) $m_T$ spectrum
to the pion (or proton) $m_T$ spectrum which we obtain by fitting the experimentally 
measured meson (or baryon) spectrum. We use pion fit function to obtain the $m_T$ spectra of mesons such as 
$\phi$, $K^{\pm}$ and $K_{s}^{0}$ and proton fit function to obtain the $m_T$ spectra of baryons such as 
$\Lambda$, $\Delta$, $\Xi$ and $\Omega$ etc.

\begin{table*}
  \caption{Particles with their mode of measurement, rapidity and $p_{T}$
    ranges for p+p collisions at different energies.} 
  
  \label{refer1}
  \begin{tabular}{|c|c|c|c|c|}
    \hline
    \cline{1-4}
    Particle    &    Mode             & $p_T$ range (GeV)   &  Rapidity range        &  Reference              \\
    
    \hline
    \multicolumn{4}{|c|} {p+p collision at $\sqrt{s}$= 62.4 GeV}                                                               \\
    \hline

    $\pi^{0}$        & $ \gamma \gamma$     & 0.6-6.7           &  $|y|<0.35$      &  PHENIX \cite{pp62pion}          \\

    $\pi^{\pm}$     &  dE/dx                & 0.3-2.85          &  $|y|<0.35$      &  PHENIX \cite{pp62chargedpionkaonproton}        \\

    $K^{\pm}$       &  dE/dx                & 0.45-1.95        &  $|y|<0.35$       & PHENIX \cite{pp62chargedpionkaonproton}     \\
       
    \hline
    \multicolumn{4}{|c|} {p +p collision at $\sqrt{s}$= 200 GeV}                                         \\
    \hline
        
    $\pi^{0}$        & $ \gamma \gamma$          & 0.6-18.9   &  $|y|<0.35$   &  PHENIX  \cite{PPG063}          \\
    
    $\pi^{\pm}$      &    dE/dx                 & 0.3-2.6     &  $|y|<0.35$    &  PHENIX  \cite{PPG030}          \\
    
    P             &      dE/dx                  & 0.468-6.5   & $|y|<0.55$    & STAR \cite{ppproton}      \\
    
    $\Lambda$            & P $ \pi^{-}$        &  0.35-4.75   & $|y|<0.75$    &   STAR \cite{ppbaryon}         \\

    $\Xi^{-}$      &   $\Lambda \pi^{-}$       &  0.67-3.37   & $|y|<0.75$     &  STAR \cite{ppbaryon}       \\
    
    & $\Sigma^{-} \gamma$      &                  &                   &             \\

    $\Omega^{-}$      & $\Lambda$ $K^{-}$     &  1.05-2.83   &  $|y|<0.75$   & STAR \cite{ppbaryon}       \\
    
    & $\Xi^{0} \pi^{-}$      &             &           &    \\ 
    
    & $\Xi^{-} \pi^{0}$      &             &           &    \\ 
        
    \hline
    \multicolumn{4}{|c|} {p+p collision at $\sqrt{s}$= 900 GeV}                                         \\
    \hline
        
    $\pi^{\pm}$      &    dE/dx                     & 0.1-2.5   &  $|y|<0.8$    & ALICE \cite{pp900pionkaonproton}          \\

    $K^{\pm}$        &    dE/dx                     & 0.2-2.3   &  $|y|<0.8$    &  ALICE  \cite{pp900pionkaonproton}        \\

    P               &    dE/dx                     & 0.3-2.3   &  $|y|<0.8$    &  ALICE \cite{pp900pionkaonproton}      \\
    
    $K_{s}^{0}$       & $\pi{+} \pi{-}$             & 0.2-2.7   &  $|y|<0.8$   & ALICE \cite{pp900strangehadron} \\

    $\Lambda$      &  P $ \pi^{-}$                  &  0.7-3.25 & $|y|<0.7$  & ALICE \cite{pp900strangehadron}         \\

    $\Xi^{-}$     &    $\Lambda \pi^{-}$           &  1.0-2.5   & $|y|<0.8$    & ALICE \cite{pp900strangehadron}         \\  
    
    & $\Sigma^{-} \gamma$      &                  &                   &             \\  
        
    \hline
  \end{tabular}
\end{table*}

\begin{table*}
  \caption{Particles with their measured decay channels, rapidity and $p_{T}$
    range for d+Au, Au+Au systems at $\sqrt{s_{NN}}$ = 200 GeV.} 
  
  \label{refer2}
  \begin{tabular}{|c|c|c|c|c|}
    \hline
    \cline{1-4}
    Particle    &    Mode                   & $p_T$ range (GeV)   &  Rapidity range        &  Reference              \\
    \hline
    \multicolumn{4}{|c|} {d+Au collision}                                         \\
    \hline
    
    P                    & dE/dx                & 0.468-3.5    & $|y|<0.75$     & STAR  \cite{dauproton}      \\
        
    $\Delta^{++}$      &    N $\pi$              &  0.3-1.5    & $|y|<0.75$     & STAR \cite{daubaryon}       \\

    \hline
    \multicolumn{4}{|c|} {Au +Au collision}                                         \\
    \hline
    
    P                &   dE/dx                  & 0.5-10.85     & $|y|<0.75$      &  STAR \cite{auauproton}      \\
       
    $\Lambda$            &   P $ \pi^{-}$      &  0.65-4.75    &   $|y|<0.75$      & STAR \cite{auaubaryon}         \\
        
    $\Xi^{-}$      &   $\Lambda \pi^{-}$       &  0.85-4.75    &   $|y|<0.75$     &  STAR  \cite{auaubaryon}       \\
    
    & $\Sigma^{-} \gamma$      &               &                   &             \\ 
        
    $\Omega^{-}$      & $\Lambda$ $K^{-}$        &  1.31-4.0    &  $|y|<0.75$     & STAR \cite{auaubaryon}       \\
    
    & $\Xi^{0} \pi^{-}$      &             &           &    \\ 
    
    & $\Xi^{-} \pi^{0}$      &             &           &    \\ 
        
    \hline
  \end{tabular}
\end{table*}

 All the mesons and baryons used in this analysis along with mode of measurement, rapidity and 
$p_{T}$ ranges with their references are listed  
in Table~\ref{refer1} for p+p and in Table~\ref{refer2} for d+Au and Au+Au systems at 
different energies. 
  The errors on the data are quadratic sums of statistical and uncorrelated systematic errors wherever available.
Here we use all data from PHENIX ($|y| < 0.35 $), STAR ($|y| < 0.5 $, $|y| < 0.75 $)  and ALICE ($|y| <0.8$)
collaborations.
 The pion and proton spectra measured in p+p collisions for different energies 
are fitted using Eq.~\ref{fitfun} and the parameters are given in Table~\ref{piscalepp} and 
\ref{protonscalepp} respectively. 
The parameters for d+Au and Au+Au collisions are given in Table~\ref{piscalenonpp} and \ref{auaucentPar} 
respectively for pions and protons.
 For Au+Au system, data corresponding to four centrality classes namely 
10-20 \%, 20-40 \%, 40-60 \% and 60-80 \% has been analyzed.

\begin{table*}
  \caption{The parameters of the Hagedorn distribution obtained by fitting pion spectra 
    measured in p+p collisions at different center of mass energies.}
  \label{piscalepp}
  \begin{tabular}{|c|c|c|c|}
    \hline
    
    Parameters for             & $\sqrt{s}$   &  $\sqrt{s}$                & $\sqrt{s}$   \\ 
    pion                     &  = 62.4 GeV        & = 200 GeV \cite{mesonscaling}  &  = 900 GeV \\
    
    \hline           
    $A$ (GeV/$c$)$^{-2}$    &  7.98 $\pm$ 0.42             &  11.11 $\pm$ 0.37        &  16.32 $\pm$ 0.81    \\
    $a$ (GeV/$c$)$^{-1}$    &  0.137 $\pm$ 0.014           &  0.32 $\pm$ 0.03         &  0.82 $\pm$ 0.17    \\
    $b$ (GeV/$c$)$^{-1}$    &  0.0 (fixed)                 &  0.024 $\pm$ 0.011       &  0.0 (fixed)        \\
    $p_0$ (GeV/$c$)         &  1.24 $\pm$ 0.026            &  0.72 $\pm$ 0.03         &  0.43 $\pm$ 0.036  \\
    $n$                     &  12.07 $\pm$ 0.11            &  8.42 $\pm$ 0.12        &  6.05 $\pm$ 0.28     \\
    \hline
  \end{tabular}
\end{table*}


\begin{table*}
  \caption{The parameters of the Hagedorn distribution obtained by fitting proton spectra 
    measured in p+p collisions at different center of mass energies.}
  \label{protonscalepp}
  \begin{tabular}{|c|c|c|c|}
    \hline
    
    Parameters for             & $\sqrt{s}$   &  $\sqrt{s}$           & $\sqrt{s}$   \\ 
    proton                           &  = 62.4 GeV    & = 200 GeV                  &  = 900 GeV \\
    
    \hline           
    $A$ (GeV/$c$)$^{-2}$         & 16.34 $\pm$ 5.847    &  168.87 $\pm$ 45.77     &  305.93  $\pm$ 38.69 \\
    $a$ (GeV/$c$)$^{-1}$         & 0.51 $\pm$ 0.108    &  0.15 $\pm$ 0.058       &  0.283  $\pm$ 0.067  \\
    $b$ (GeV/$c$)$^{-1}$         & 0.0 (fixed)        &  0.0 (fixed)            &  0.0 (fixed)       \\
    $p_0$ (GeV/$c$)              & 0.87 $\pm$ 0.104    &  0.69 $\pm$ 0.029        &  0.36 $\pm$ 0.007   \\
    $n$                          & 11.49 $\pm$ 0.77    &  9.86 $\pm$ 0.13        &  7.17 $\pm$ 0.005    \\
    \hline
  \end{tabular}
\end{table*}

\begin{table}
  
  \caption{The parameters of the Hagedorn distribution obtained by fitting pion spectra 
    measured in d+Au and Au+Au (10-20\%) collisions at $\sqrt{s_{NN}}$ = 200 GeV.}
  \label{piscalenonpp}
  \begin{tabular}{|c|c|c|}
    \hline
    Parameters for             &  d+Au system         & Au+Au system (10-20\% cent.)   \\ 
    pion                     &    \cite{mesonscaling} & \\
    
    \hline           
    $A$ (GeV/$c$)$^{-2}$     &  52.30 $\pm$ 1.63  &    1564.36 $\pm$ 36.68   \\
    $a$ (GeV/$c$)$^{-1}$     &  0.24 $\pm$ 0.01   &    0.438 $\pm$ 0.007      \\
    $b$ (GeV/$c$)$^{-1}$     &  0.11 $\pm$ 0.01  &    0.213 $\pm$ 0.007       \\
    $p_0$ (GeV/$c$)          &  0.77 $\pm$ 0.02   &    0.713 $\pm$ 0.004      \\
    $n$                       &  8.46 $\pm$ 0.07   &    8.383 $\pm$ 0.017     \\
    \hline
  \end{tabular}
\end{table}


\begin{table*}
  \caption{The parameters of the Hagedorn distribution obtained by fitting proton spectra 
    measured in d+Au (MB) and Au+Au collisions (for different centralities) at $\sqrt{s_{NN}}$ = 200 GeV.}
  \label{auaucentPar}
  \begin{tabular}{|c|c|c|c|c|c|}
    \hline
                               & d+Au collision           &  \multicolumn{4}{|c|} {Au+Au collision for different centralities}              \\
    \cline{3-6}
    Parameters for proton              &            &    10-20 \%               &  20-40 \%              &   40-60 \%        &  60-80 \%  \\
    \hline           
    $A$ (GeV/$c$)$^{-2}$    &  0.35 $\pm$ 0.10    & 0.3 $\pm$ 0.2          & 0.98 $\pm$ 0.13    & 0.43 $\pm$ 0.13     &  0.90 $\pm$ 0.34   \\
    $a$ (GeV/$c$)$^{-1}$    &  1.63 $\pm$ 0.17     & 1.07 $\pm$ 0.08      & 0.93 $\pm$ 0.16    & 1.12 $\pm$ 0.054    & 1.16 $\pm$ 0.09 \\    
    $b$ (GeV/$c$)$^{-2}$    &  0.0(fixed)         & 0.0(fixed)           &   0.0(fixed)             &   0.0(fixed)          & 0.0(fixed)   \\
    $p_0$ (GeV/$c$)         &  1.09 $\pm$ 0.045      & 2.3 $\pm$ 0.12              & 1.95 $\pm$ 0.18    & 1.78 $\pm$ 0.051    & 1.36 $\pm$ 0.05 \\
    $n$                     &  7.06 $\pm$ 0.24      & 10.22 $\pm$ 0.21        & 10.34 $\pm$ 0.35   & 9.48  $\pm$ 0.16    & 9.05 $\pm$ 0.19     \\
    \hline
  \end{tabular}
\end{table*}

\section{Results and Discussions}
\begin{figure*}
  \includegraphics[width=0.9\textwidth]{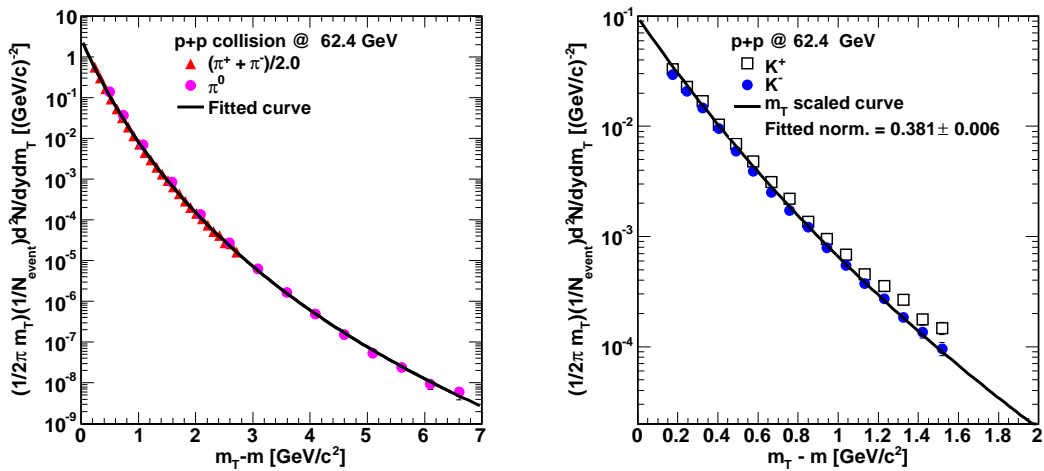}
  \caption{(Color online) The invariant yield of (a) $\pi^{\pm}$ \cite{pp62chargedpionkaonproton}, $\pi^{0}$ \cite{pp62pion} 
    and (b) K$^{\pm}$ \cite{pp62chargedpionkaonproton} 
    as a function of $m_{T}$ for p+p collision at $\sqrt s$ = 62.4 GeV.}
  \label{ppmeson}
  
\end{figure*}

Figure~(\ref{ppmeson}a) shows the invariant yield of $\pi^{\pm}$ \cite{pp62chargedpionkaonproton}, $\pi^{0}$ \cite{pp62pion}  
and (\ref{ppmeson}b) shows the invariant yield of K$^{\pm}$ \cite{pp62chargedpionkaonproton} as a function of $m_T$ measured 
in p+p collisions at $\sqrt{s}$ = 62.4 GeV.
Here the data of neutral and charged pion are fitted with modified Hagedorn function shown in solid black curve. 
The corresponding fit parameters of pion are given in Table \ref{piscalepp}.  
In case of kaon, the solid line is obtained from pion fit function using $m_T$ scaling; the relative normalization has been used to fit
the kaon spectra. There is difference in the measured data for K$^+$ and K$^-$ and the $m_T$ scaled curve
reproduces only K$^-$ well. The relative normalization for kaons are given in Table \ref{scaledmesons} .




\begin{figure*}
  \includegraphics[width=0.9\textwidth]{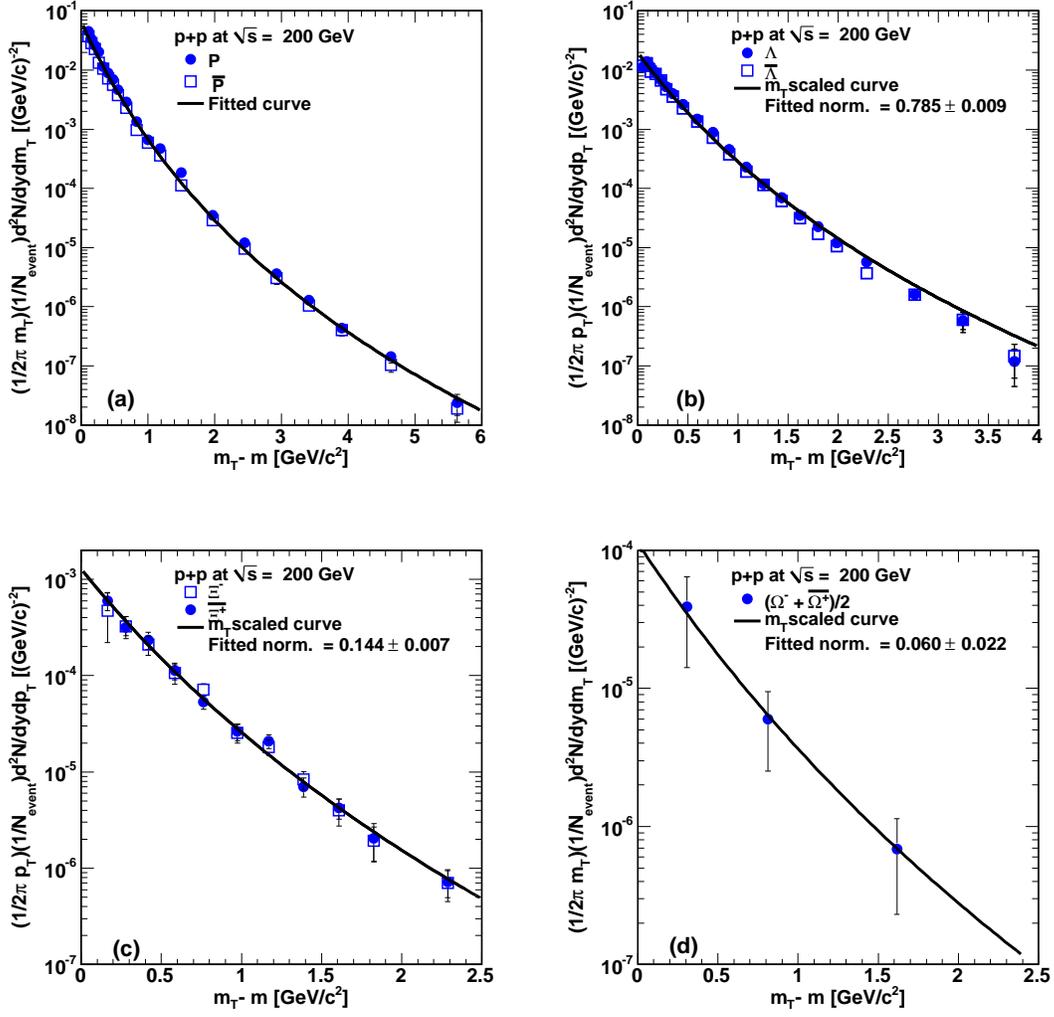}
  \caption{(Color online) The invariant yield of (a) proton (solid circle), anti-proton (open square) \cite{ppproton}, 
    (b) $\Lambda$ (solid circle), $\bar{\Lambda}$ (open square) \cite{ppbaryon},
    (c) $\Xi^{-}$ (solid circle), $\bar{\Xi^{+}}$ (open square) \cite{ppbaryon} and 
    (d) $(\Omega^{-} + \bar{\Omega^{+}})/2$ (solid circle) \cite{ppbaryon} 
    as a function of $m_{T}$ for p+p collision at $\sqrt s$ = 200 GeV.}
  \label{ppbaryon}
  
\end{figure*}

\begin{figure*}
  \includegraphics[width=0.9\textwidth]{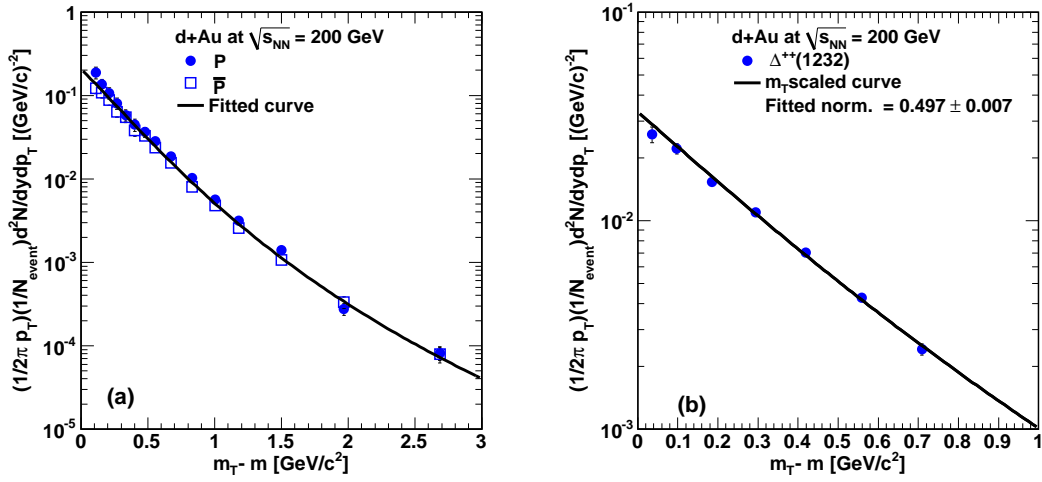}
  \caption{(Color online) The invariant yield of (a) proton (solid circle), anti-proton (open square) \cite{dauproton}  and 
    (b) $\Delta^{++}$ baryons \cite{daubaryon} as a function of $m_{T}$ for d+Au collision at $\sqrt{s_{NN}}$ = 200 GeV.}
  \label{daubaryon}
\end{figure*}

\begin{figure*}
  \includegraphics[width=0.61\textwidth]{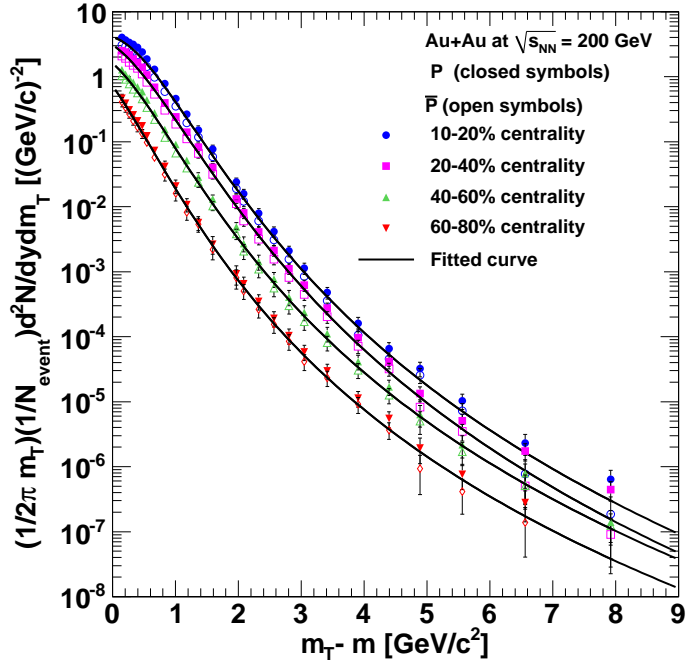}
  \caption{(Color online) The invariant yield of proton (closed symbols) and anti-proton (open symbols) \cite{auauproton}
    as a function of $m_{T}$ 
    in Au+Au collisions at $\sqrt{s_{NN}}$ = 200 GeV for different centralities. The solid lines are Hagedorn fit function.}
  \label{auauproton}
  
\end{figure*}

\begin{figure*}
  \includegraphics[width=0.61\textwidth]{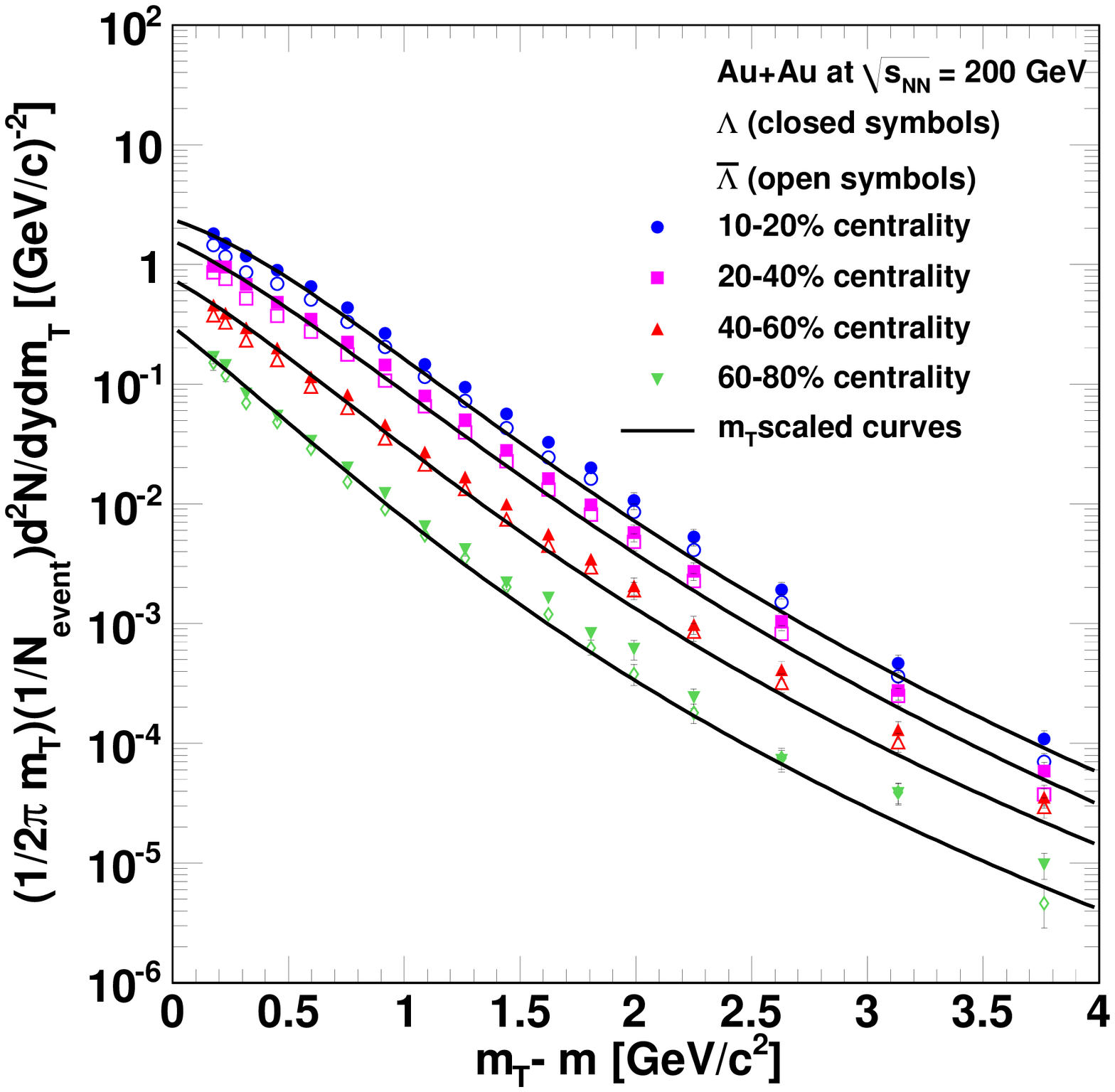}
  \caption{(Color online) The invariant yield of $\Lambda$ (closed symbols) and $\bar{\Lambda}$ (open symbols) \cite{auaubaryon} 
    as a function of $m_{T}$ in Au+Au collisions at $\sqrt{s_{NN}}$ = 200 GeV for different centralities.
    The solid lines are obtained using $m_{T}$ scaling; the relative normalization has been 
    used to fit the measured spectra.}
  \label{auauLambda}
\end{figure*}

\begin{figure*}
  \includegraphics[width=0.61\textwidth]{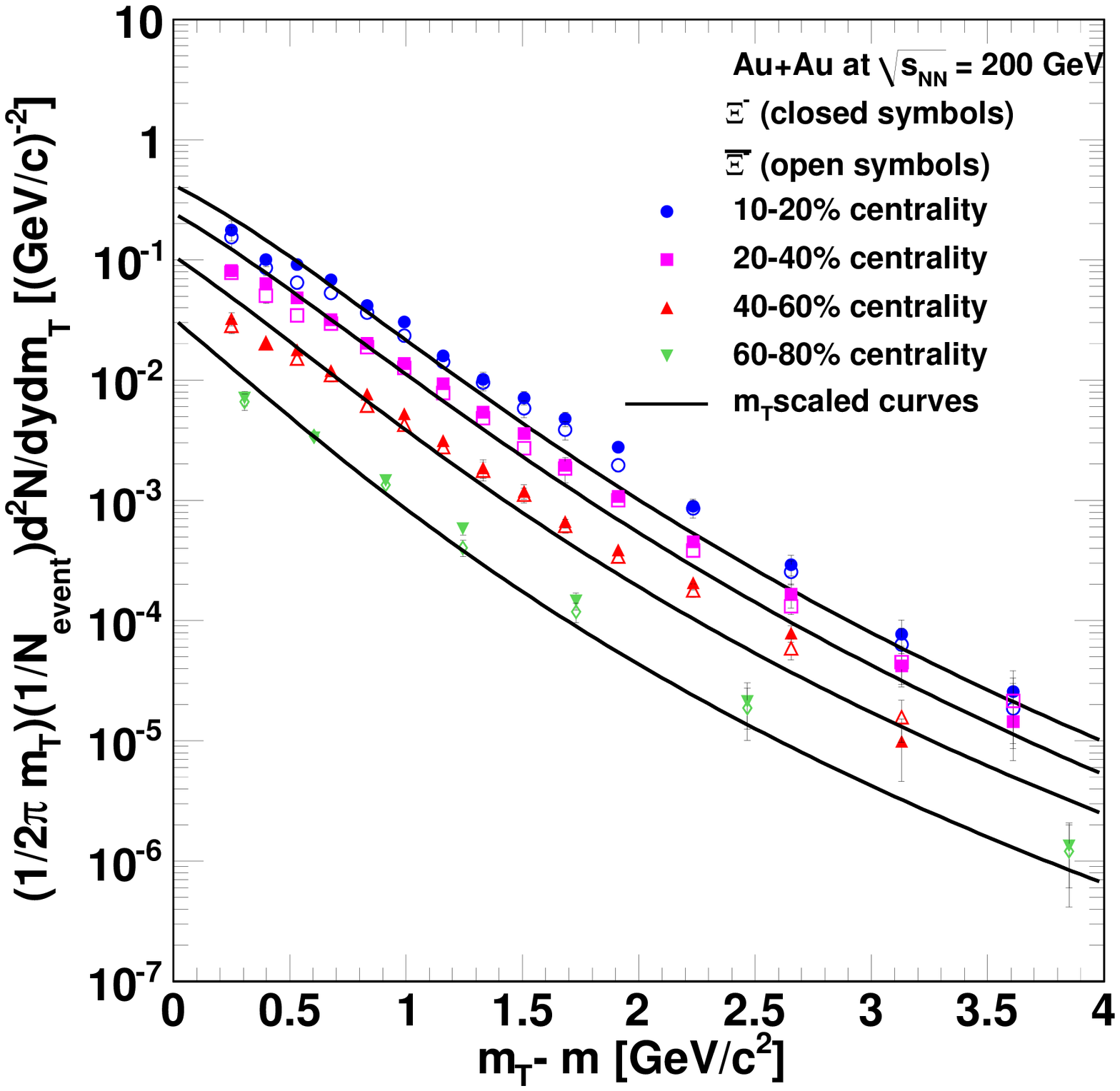}
  \caption{(Color online) The invariant yield of $\Xi^{-}$ (closed symbols) and $\bar{\Xi^{+}}$ (open symbols)  \cite{auaubaryon}
    as a function of $m_{T}$ in Au+Au collisions at 
    $\sqrt{s_{NN}}$ = 200 GeV for different centralities. The solid lines are obtained using $m_{T}$ scaling; 
    the relative normalization has been used to fit the measured spectra.}
  \label{auauXi}
  
\end{figure*}

\begin{figure*}
  \includegraphics[width=0.61\textwidth]{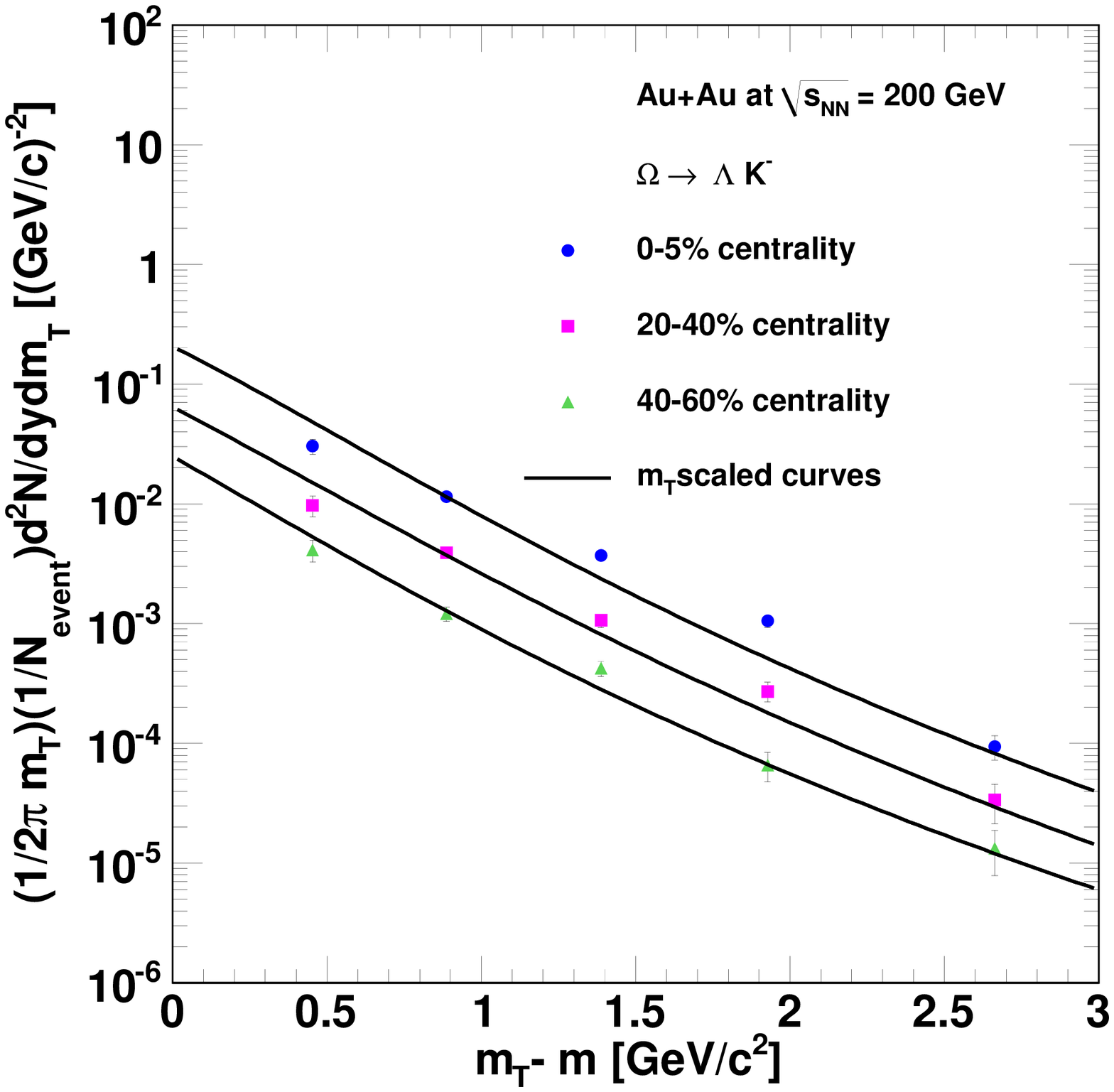}
  \caption{(Color online) The invariant yield of $\Omega$ (closed symbols) \cite{auaubaryon} as a function of $m_{T}$
    in Au+Au collisions at
    $\sqrt{s_{NN}}$ = 200 GeV  for different centralities. The solid lines are obtained using $m_{T}$ scaling; 
    The Relative normalization has been used to fit the measured spectra. }
  \label{auauOmega}
  
\end{figure*}



\begin{figure*}
  \includegraphics[width=0.9\textwidth]{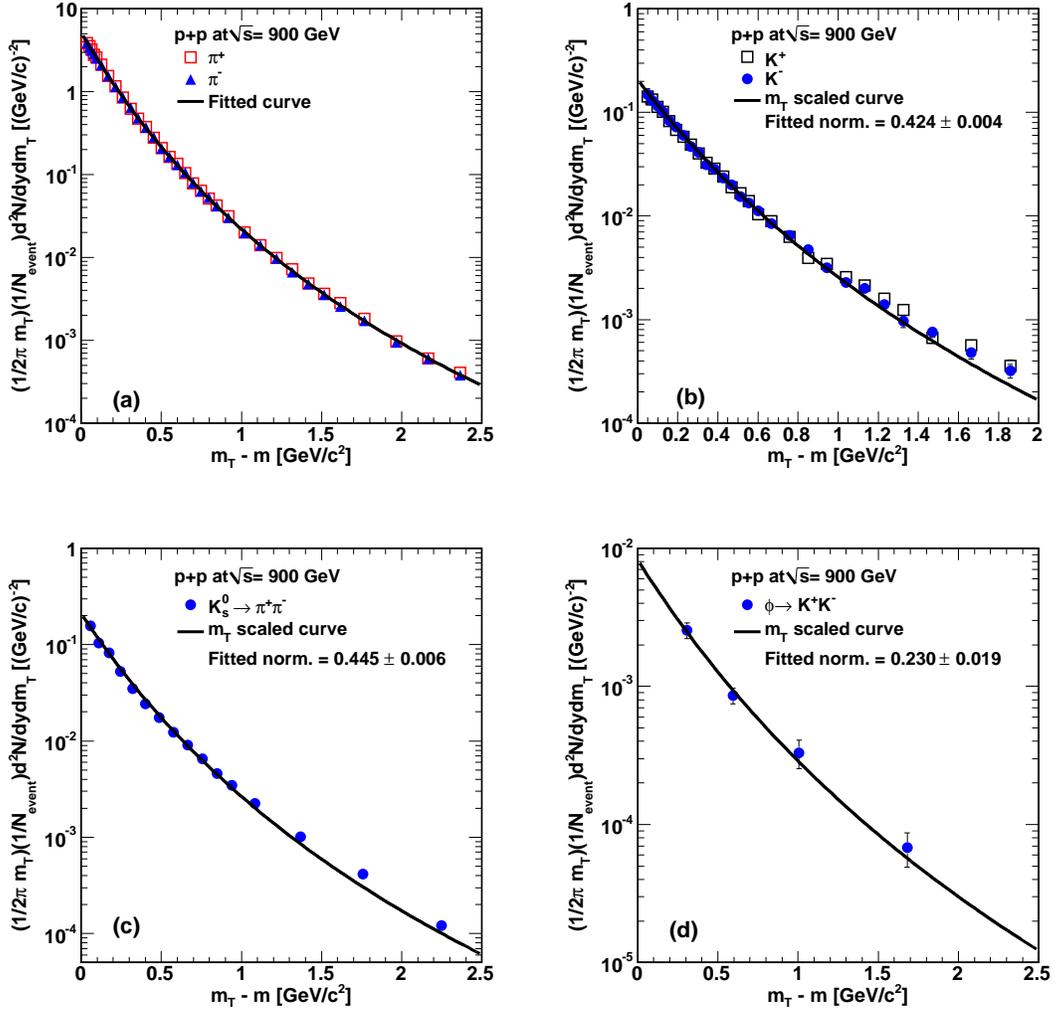}
  \caption{(Color online) The invariant yield of (a) $\pi^{\pm}$ (closed triangle) \cite{pp900pionkaonproton}, 
    (b) K$^{\pm}$ (open square, solid circle) \cite{pp900pionkaonproton},
    (c) $K_{S}^{0}$ (solid circle)\cite{pp900strangehadron} and 
    (d) $\phi$ (solid circle) \cite{pp900strangehadron} as a function of $m_{T}$ for p+p collision at 
    $\sqrt s$ = 900 GeV. The solid lines are obtained using $m_{T}$ scaling.}
  \label{ppmesonlhc}
  
\end{figure*}

\begin{figure*}
  \includegraphics[width=0.9\textwidth]{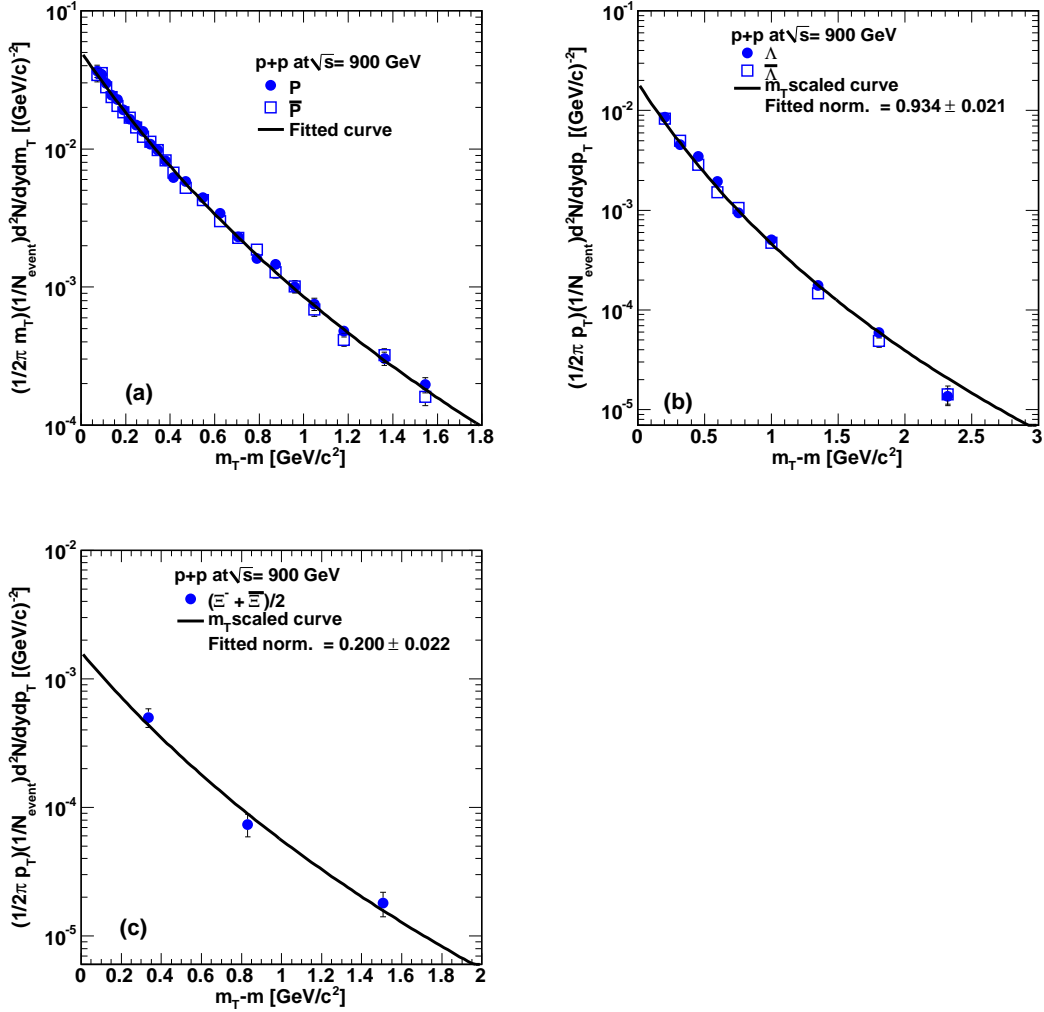}
  \caption{(Color online) The invariant yield of (a) proton (solid circle), anti-proton (open square) \cite{pp900pionkaonproton}, 
    (b) $\Lambda$ (solid circle), $\bar{\Lambda}$ (open square) \cite{pp900strangehadron} and 
    (c) $\Xi$ (solid circle) \cite{pp900strangehadron} as a function of $m_{T}$  for p+p collision at $\sqrt s$ = 900 GeV.
    The solid lines are obtained using $m_{T}$ scaling.}
  \label{ppbaryonlhc}
  
\end{figure*}


\begin{figure*}
  \includegraphics[width=0.9\textwidth]{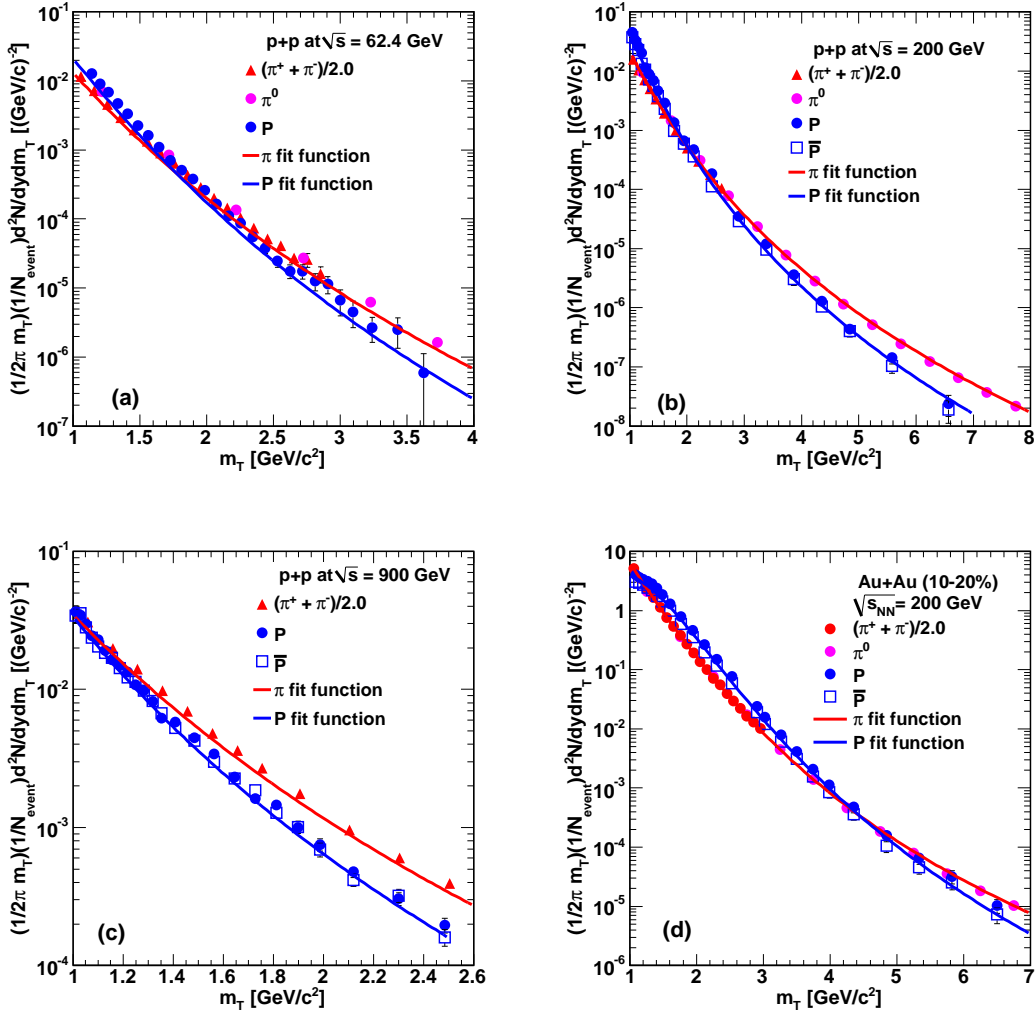}
  \caption{(Color online) The invariant yields of proton, anti-proton, $\pi^{\pm}$ and $\pi^{0}$ 
as a function of $m_{T}$ for 
 (a)  p+p at $\sqrt{s}$ = 62.4 GeV \cite{pp62chargedpionkaonproton, pp62pion}, 
 (b)  p+p at $\sqrt{s}$ = 200 GeV \cite{ppproton,PPG030, PPG063},
 (c)  p+p at $\sqrt{s}$ = 900 GeV \cite{pp900pionkaonproton},
 (d) Au+Au (for 10-20\% centrality) at  $\sqrt{s_{NN}}$ = 200 GeV \cite{auauproton, PPG026, PPG080}.
    The solid lines show the modified Hagedorn fit function drawn over proton and pion. }

  \label{mtscaling_protonpion_ppauau}
  
\end{figure*}


\begin{figure}
  \includegraphics[width=0.61\textwidth]{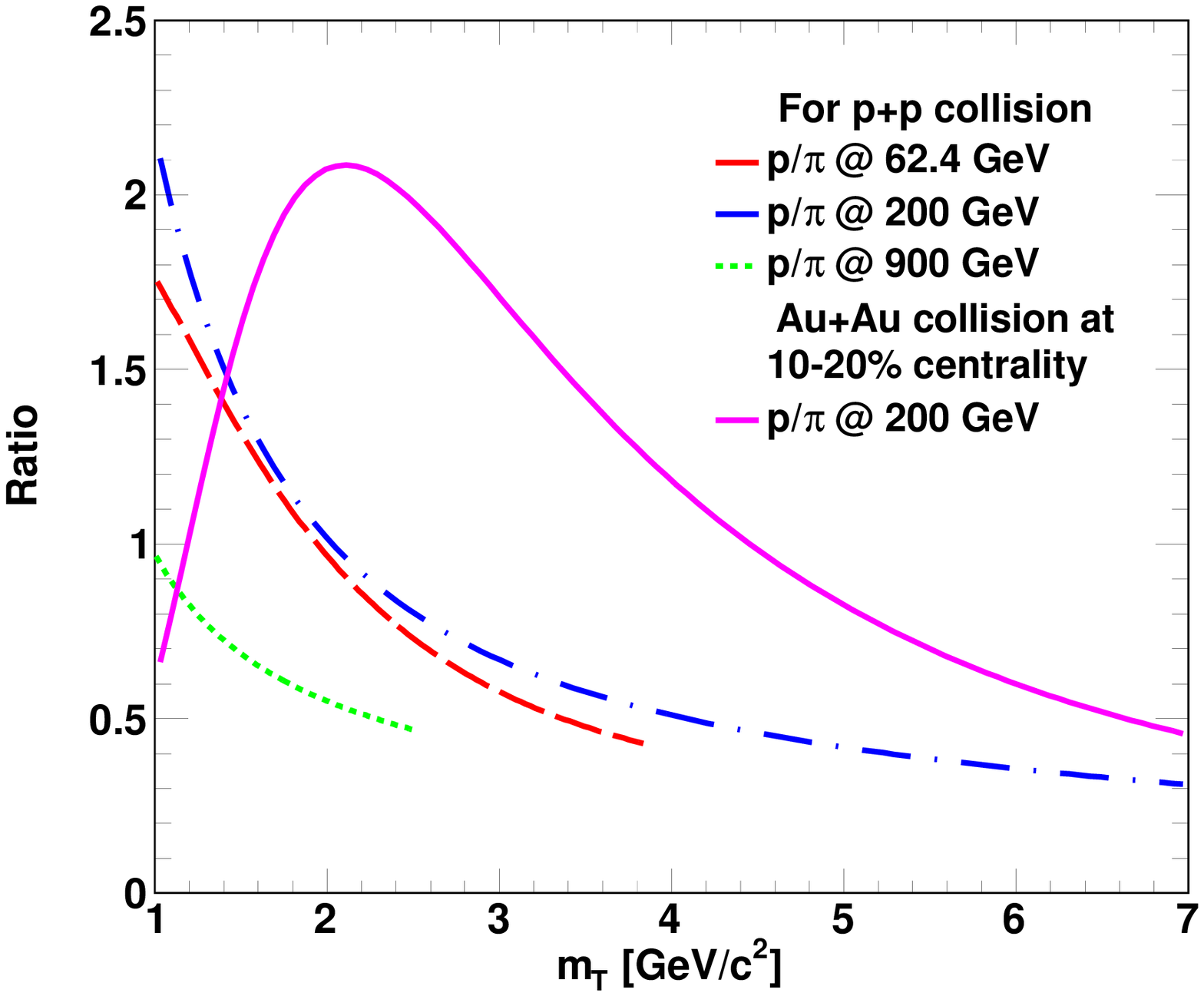}
  \caption{(Color online) The ratio of proton fit function to the pion fit function (p/$\pi$) as a function of $m_{T}$
    for p+p collisions at different energies and Au+Au collisions (10-20\% centrality) at 200 GeV.}
  \label{Ratio}
  
\end{figure}



Figure~\ref{ppbaryon} shows the invariant yield of different baryons as a function of $m_{T}$ 
in p+p collisions at $\sqrt{s}$ = 200 GeV. 
Here the data of proton and anti-proton \cite{ppproton} are fitted with the modified Hagedorn 
function (shown in solid black curve). The fit parameters are given in Table~\ref{auaucentPar}
which are used to get the $m_{T}$ scaled 
spectra for measured baryons such as $\Lambda$, $\bar{\Lambda}$,
$\Xi^{-}$, $\bar{\Xi^{+}}$ and $\Omega^-$, $\bar{\Omega^+}$  \cite{ppbaryon} as shown in figure.  
The relative normalizations of baryon (given in Table~\ref{baryonratio}) to proton
have been obtained by fitting  the measured spectra. 
 It is seen that the $m_{T}$ spectrum of all the baryons are well described by the 
$m_T$ scaled curve. In case of $\Lambda$ there are deviations of data from the curve in high $p_T$ region.


The Figure~(\ref{daubaryon}a) shows the invariant yield of proton 
and anti-proton \cite{dauproton} in d+Au system at $\sqrt{s_{NN}}$ = 200 GeV as a function 
of $m_T$. The solid black curve is fit function (Eq.~\ref{fitfun}) the parameters of which
are given in the Table \ref{auaucentPar}.
The Figure~(\ref{daubaryon}b) shows the invariant yield  of $\Delta^{++}$ \cite{daubaryon} 
as a function of $m_T$. The solid line is $m_T$ scaled curve; the relative normalisation given in Table \ref{baryonratio}
is obtained by fitting the measured data reproducing the shape of the measured spectrum very well.

Figure~\ref{auauproton} shows the invariant yield of proton and anti-proton \cite{dauproton} as a function of $m_{T}$ for 
Au+Au collsion at $\sqrt{s_{NN}}$ = 200 GeV for different centralities.
The solid black curves are the fitted modified Hagedorn functions with parameters given in Table~\ref{auaucentPar}. 
Figure~\ref{auauLambda} shows the invariant yield of $\Lambda$ and  $\bar{\Lambda}$ \cite{auaubaryon} for different centralities
 as a function of $m_{T}$ for Au+Au collision at $\sqrt{s_{NN}}$ = 200 GeV.
The solid black curves are the $m_T$ scaled curves obtained from the proton spectra of 
corresponding centralities with a fitted normalization given in Table~\ref{baryonratio}. 
Figure~\ref{auauXi} shows the invariant yield of $\Xi$ and $\bar{\Xi}$ \cite{auaubaryon} for different 
centralities as a function of $m_{T}$
for Au+Au collision at $\sqrt{s_{NN}}$ = 200 GeV.
The solid black curves are the $m_T$ scaled curves obtained from the proton spectra of 
corresponding centralities with a fitted normalization given in Table~\ref{baryonratio}. 
 Here we see that, in both the cases for $\Lambda$, $\bar{\Lambda}$ and  $\Xi^-$, $\bar{\Xi^+}$, 
the experimental data are in good agreement with the $m_T$ scaled 
curve for the most peripheral centrality bin.
  But for other centralities, the nature of the $m_T$ scaled curves
are different  from the shape of data. 
 Same obeservation is made from Figure~\ref{auauOmega} which shows the 
invariant yield of $\Omega$ \cite{auaubaryon} 
as a function of $m_{T}$ for  Au+Au collision at $\sqrt{s_{NN}}$ = 200 GeV.

 Figure~(\ref{ppmesonlhc}a) shows the invariant yield of $\pi^{\pm}$ \cite{pp900pionkaonproton}
as a function of $m_T$ for p+p collision at $\sqrt{s}$ = 900 GeV. The solid black curve is 
fit function (Eq.~\ref{fitfun}) whose parameters are given in the Table \ref{piscalepp}.
 Figures~\ref{ppmesonlhc} (b), (c) and (d) show the invariant yields of $K^{\pm}$ \cite{pp900pionkaonproton}, 
$K_S^0$ \cite{pp900strangehadron}, and $\phi$ \cite{pp900strangehadron}, respectively 
as a function of $m_T$. 
 The solid black curves are the $m_T$ scaled curves obtained from the pion spectrum with a fitted 
normalization given in Table~\ref{scaledmesons}.
From this  we clearly see that the $m_T$ scaled curves do not reproduce mesons
for p+p at 900 GeV that well.

The Figure~\ref{ppbaryonlhc} shows the invariant yields of proton and anti-proton \cite{pp900pionkaonproton} 
as a function of $m_T$ for p+p collision at $\sqrt{s}$  = 900 GeV. 
The solid black curve is the fitted modified Hagedorn function with parameters given in Table~\ref{protonscalepp}. 
 The Figures~\ref{ppbaryonlhc} (b) and (c) show the invariant yields of $\Lambda$ \cite{pp900strangehadron}
 and $\Xi$ \cite{pp900strangehadron}.
 The solid black curves are the $m_T$ scaled curves obtained from the proton spectra 
with fitted normalizations given in the Tabe \ref{baryonratiopp}.
Here the $m_T$ scaled curve is in good agreement with $\Lambda$ and with small data set of $\Xi$ at $\sqrt{s}$ = 900 GeV.

 Figure~\ref{mtscaling_protonpion_ppauau} shows the invariant yields of proton (P), anti-proton ($\bar{P}$) 
and pion ($\pi$) as a function of $m_{T}$ for p+p collisions at $\sqrt{s}$  = 62.4, 200 and 900 GeV and 
for Au+Au collisions (10-20\% centrality) at $\sqrt{s_{NN}}$  = 200 GeV along with their fit functions
obtained from Eq.~\ref{fitfun}. 
 The shapes of pions and protons $m_T$ spectra are different for all systems at all energies.
 In comparision to the proton and pion production at low energies (e.g. 62.4 and 200 GeV),
 the proton production is less as compared to the pion production at 900 GeV.

Figure~\ref{Ratio} shows the ratio of proton to pion (p/$\pi$) (obtained from the 
fit functions in  Figure~\ref{mtscaling_protonpion_ppauau}) vs $m_{T}$  for p+p collisions 
at $\sqrt{s}$  = 62.4, 200 and 900 GeV and 
for Au+Au collisions (10-20\% centrality) at $\sqrt{s_{NN}}$  = 200 GeV.
 It is seen that for p+p collisions, the shape of p/$\pi$ curves are 
same at 62.4, 200 and also at 900 GeV, but the ratio of p/$\pi$ is very small at $\sqrt{s_{NN}}$  = 900 GeV.
 The enhancement of p/$\pi$ ratio at intermediate $m_T$ in Au+Au is understood 
in terms of recombination model \cite{recommodel}.

\begin{table*}
  \caption{The relative normalization (S) of mesons obtained by fitting the $m_{T}$ scaled spectra 
    with the measured spectra for p+p collisions at $\sqrt s $ = 62.4, 200 and 900 GeV.}
  \label{scaledmesons}
  \begin{tabular}{|c|c|c|c|c|c|}
    \hline
    
    S                  & $\sqrt s $ = 62.4 GeV &  $\sqrt s $ = 200 GeV     &   $\sqrt s $ = 900 GeV                  \\
                       &                       & From \cite{mesonscaling}  &                    \\   
    \hline 
    $K$/$\pi$           & 0.38 $\pm$ 0.02          & 0.422 $\pm$ 0.014                 & 0.424 $\pm$ 0.02       \\
    $K_{S}^0$/$\pi$     &                          &                                    & 0.44 $\pm$ 0.02       \\
    
    $\phi$/$\pi$        &                          &  0.233 $\pm$ 0.013                 & 0.23 $\pm$ 0.02 \\
    
    \hline
  \end{tabular}
\end{table*}

\begin{table}
\caption{The relative normalization (S) of baryons obtained by fitting the $m_{T}$ scaled spectra 
      with the measured spectra for p+p collisions at different center of mass energies.}
\label{baryonratiopp}
\begin{tabular}{|c|c|c|}
\hline
      S                   &  $\sqrt{s}$           & $\sqrt{s}$    \\ 
                          & = 200 GeV                 &  = 900 GeV          \\
\hline           
    $\Lambda$/p           &  0.785 $\pm$ 0.021        &    0.93 $\pm$ 0.2   \\
    $\Xi$/p               &  0.14 $\pm$ 0.03          &    0.2 $\pm$ 0.02  \\
    $\Omega$/p            & 0.060 $\pm$ 0.022         &                     \\
\hline
\end{tabular}
\end{table}

\begin{table*}
  \caption{The relative normalization (S) of baryons obtained by fitting the $m_T$ scaled spectra
    with the measured spectra for different collision systems at $\sqrt{s_{NN}} $ = 200 GeV.}
  \label{baryonratio}
  \begin{tabular}{|c|c|c|c|c|c|c|}
    \hline
    & p+p collision    &   d+Au collision  &  \multicolumn{4}{|c|} {Au+Au collision for different centralities}              \\
    \cline{4-7}
    S              &                   &                 &  10-20\%            &  20-40 \%             &   40-60 \%            &  60-80 \%          \\
    \hline 
    $\Lambda$/p     & 0.785 $\pm$ 0.021    &                 & 0.70 $\pm$ 0.09     & 0.70 $\pm$ 0.09     &0.7 $\pm$ 0.2        & 0.75 $\pm$ 0.28  \\
    
    $\Xi$/p         & 0.140 $\pm$ 0.030    &                 & 0.18 $\pm$ 0.02      & 0.18 $\pm$ 0.02     &0.18 $\pm$ 0.05      & 0.17 $\pm$ 0.06  \\
    
    $\Omega$/p  & 0.060 $\pm$ 0.022    &                & 0.16 $\pm$ 0.008(0-5\%) &0.13 $\pm$ 0.01     &0.12 $\pm$ 0.03       &                     \\
    
    $\Delta$/p  &                   & 0.497  $\pm$ 0.016  &                         &                    &                     & \\
    
    \hline
  \end{tabular}
\end{table*}

\section{conclusion}
  In summary, we have made a systematic study of transverse mass spectra of mesons and baryons at 
RHIC and LHC energies. 
  For p+p and d+Au collision at $\sqrt{s_{NN}}$ = 200 GeV
the baryon $m_T$ spectra scale with proton $m_T$ spectrum just like all meson
spectra scale with pion spectra. In p+p 200 GeV, $m_T$ scaling reproduce data well for all the 
baryons except $\Lambda$ as there is a disagreement at high $p_T$ points. 
  In case of Au+Au collisions, the strange baryons behave differently from protons for all centralities
but there is better agreement in the two for peripheral collisions.
  This study has also been performed on the meson and baryon spectra in p+p collisions
at 62.4 GeV which lies in between the highest RHIC energy and SPS energy where the $m_T$ 
scaling was first observed and we find that the $m_T$ scaling of 
mesons work at 62.4 GeV (Baryon data at 62.4 GeV is not available). 
 We test the $m_T$ scaling at LHC with the first meson and baryon spectra measured in p+p collisions at  
900 GeV and find that the $m_T$ spectra of kaons and $\phi$ do not scale with pions that well. 
  At 900 GeV, $\Lambda$ is produced well
by $m_T$ scaling where as the $\Xi$ spectrum is reproduced well within the limited availability of data points. 
 The shapes of pions and protons $m_T$ spectra are different at all energies.
Proton to pion ratio is studied at different energies (using fit functions) and for different systems 
and is found that for p+p collisions, the behaviour of p/$\pi$ curves are similar at 62.4, 200 and also
at 900 GeV.
  It will be interesting to have such study at higher energies such as 2.76 TeV and 7 TeV, once the data 
tables are available. More comprehensive measurements at lower energies such as 62.4 GeV will benifit
to draw a systematic picture of hadron production as a function of energy.

\section{Acknowledgements}
  We acknowledge the financial support from Board of Research in Nuclear 
Sciences (BRNS) for this project. We thank S. Kailas, A. Chatterjee and A. K. Mohanty for fruitful
discussions.


\begin{thebibliography}{50}

\bibitem{WP} I. Arsene et. al. (BRAHMS), Nucl. Phys. A{\bf 757}, 1 (2005); 
 B. B. Back et. al. (PHOBOS), Nucl. Phys. A{\bf 757}, 28 (2005); 
 J. Adams et. al. (STAR), Nucl. Phys. A{\bf 757}, 10 (2005); 
 K. Adcox et. al. (PHENIX), Nucl. Phys. A{\bf 757}, 184 (2005). 






\bibitem{HADPHEN} S. S. Adler et al. (PHENIX), Phys. Rev. Lett. {\bf 94}, 082302 (2005).

\bibitem{PHOPHEN} A. Adare et al. (PHENIX), Phys. Rev. Lett. {\bf 104}, 132301 (2010). 
  
\bibitem{DIELEC} A. Adare et al. (PHENIX), Phys. Rev. C{\bf 81}, 034911 (2010). 

\bibitem{CMSJET}  Chatrchyan S. et al. (CMS), Phys. Rev. C{\bf 84}, 024906 (2011).


\bibitem{CMSUP} Chatrachyan S. et al. (CMS) Phys. Rev. Lett. {\bf 107}, 052302 (2011).

\bibitem{CMSQUARK} Chatrchyan S. et al. (CMS) arXiv:1201.5069 [nucl-ex] (2012).

\bibitem{LHC} Chatrchyan S. et al. (CMS) JHEP 1108  086 (2011). 



\bibitem{WA80} R. Albrecht et. al. (WA80 collaboration), Phys. Lett. B{\bf 361}, 14 (1995);
                       arXiv:hep-ex/9507009.

\bibitem{COCK} A. Adare et al. (PHENIX), Phys. Lett.B{\bf 670}, 313 (2009).  
  
\bibitem{mesonscaling} P. K. Khandai, P. Shukla and V. Singh, Phys. Rev. C{\bf  84}, 054904 (2011).








\bibitem{PPG077} A. Adare et al. (PHENIX), arXiv:1005.1627(2010).

\bibitem{PPG099} A. Adare et al. (PHENIX), Phys.Rev.D{\bf 83}, 052004 (2011).

\bibitem{SEANA} S. Kelly (PHENIX), J. Phys. G{\bf 30},  S1189 (2004).




 
\bibitem{RATPHI} K. Adcox et al. (PHENIX), Phys. Rev. Lett {\bf 88}, 192303 (2002). 

 \bibitem{pp62chargedpionkaonproton} A. Adare et al. (PHENIX), Phys. Rev. C{\bf 83}, 064903 (2011).

\bibitem{pp62pion} A. Adare et al. (PHENIX)  Phys. Rev. D{\bf 79}, 012003 (2009).

\bibitem{ppproton} B. I. Abelev et al. (STAR), Phys. Rev. C{\bf 75}, 64901 (2007).  

\bibitem{ppbaryon} J. Adams et al. (STAR), Phys. Lett. B{\bf 637}, 161 (2006).  

\bibitem{dauproton} J. Adams et al. (STAR), Phys. Lett. B{\bf 616}, 8 (2005).

\bibitem{daubaryon} J. Adams et al. (STAR), Phys. Rev. C{\bf 78}, 44906 (2008).

\bibitem{auauproton} J. Adams et al. (STAR), Phys. Rev. Lett. {\bf 97}, 132301 (2006).

\bibitem{auaubaryon} J. Adams et al. (STAR), Phys. Rev. Lett. {\bf 98}, 62301 (2007).

\bibitem{pp900pionkaonproton} K. Aamodt et al. (ALICE) Eur. Phys. J. C{\bf 71}(6), 1655 (2011).

\bibitem{pp900strangehadron} K. Aamodt et al. (ALICE) Eur. Phys. J. C{\bf 71}(3), 1594 (2011).


\bibitem{PPG063} A. Adare et al. (PHENIX), Phys. Rev. D{\bf 76}, 051106(R) (2007).


\bibitem{PPG030} S. S. Adler et al. (PHENIX), Phys. Rev. C{\bf 74}, 024904 (2006). 


\bibitem{PPG080} A. Adare et al. (PHENIX), Phys. Rev. Lett. {\bf 101}, 232301 (2008).

 \bibitem{PPG026} S. S. Adler et al. (PHENIX), Phys. Rev. C{\bf 69}, 034909 (2004).

\bibitem{recommodel} R. Fries, V. Greco and P. Sorensen, arXiv:0807.4939v1 [nucl-th] (2008).

\end{thebibliography}
\end{document}